\documentstyle[12pt,fleqn]{article}

\def\soc{{\rm C}_{60}}
\def\rug{{\rm C}_{70}}
\def\beeq{\begin{equation}}
\def\eneq{\end{equation}}
\def\beeqa{\begin{eqnarray}}
\def\eneqa{\end{eqnarray}}

\def\eps{\varepsilon}

\addtocounter{section}{-1}
\begin{document}

\begin{center}

{\large {\bf{Optical absorption spectra\\
in fullerenes C$_{\rm {\bf 60}}$ and C$_{\rm {\bf 70}}$:\\
Effects of Coulomb interactions,\\
lattice fluctuations, and anisotropy\\
} } }

\vspace{1cm}

{\rm Kikuo Harigaya$^*$ and Shuji Abe}\\

\vspace{1cm}

{\sl Fundamental Physics Section, Physical Science Division,\\
Electrotechnical Laboratory,\\
Umezono 1-1-4, Tsukuba, Ibaraki 305, Japan}
\vspace{1cm}

(Received~~~~~~~~~~~~~~~~~~~~~~~~~~~~~~~~~~~)
\end{center}


\Roman{table}

\vspace{1cm}

\noindent
{\bf ABSTRACT}

\noindent
Effects of Coulomb interactions and lattice fluctuations in the optical
absorption spectra of $\soc$ and $\rug$ are theoretically investigated
by using a tight binding model with long-range Coulomb interaction and
bond disorder.  Anisotropy effects in $\rug$ are also considered.
Optical spectra are calculated by using the Hartree-Fock approximation
followed by the configuration interaction method.  The main conclusions
are as follows:
(1)  The broad peaks at excitation energies, 3.7eV, 4.7eV, and 5.7eV,
observed in experiments of $\soc$ molecules in a solution are reasonably
described by the present theory.  Peak positions and relative oscillator
strengths are in overall agreement with the experiments.  The broadening
of peaks by lattice fluctuations is well simulated by the bond disorder model.
(2)  The optical gap of $\rug$ is larger when the electric field
of light is parallel to the long axis of the molecule.
The shape of the frequency
dispersion also depends on the orientation of the molecule.
These properties are common in the free electron model and the model
with Coulomb interactions.
(3) The spectrum of $\rug$ averaged over bond disorder and random
orientations is compared with experiments in a solution.
There is an overall agreement about the spectral shape.
Differences in the spectra of $\soc$ and $\rug$ are discussed
in connection with the symmetry reduction from a soccerball to a rugbyball.

\mbox{}


\pagebreak


\section{INTRODUCTION}

Recently, fullerenes C$_N$ ($N = 60, 70, 76$, and so on) with
hollow cage structures have been intensively investigated.
Many optical experiments have been performed, and interesting
properties originating from $\pi$ electrons delocalized on
molecule surfaces have been revealed.  They include the optical
absorption spectra of $\soc$ [1,2] and $\rug$ [1], and the large optical
nonlinearity of $\soc$ [3,4].  The nonlinearity in the third harmonic
generation (THG) is of the order $10^{-11}$esu, and this
largeness is attractive in the scientific as well as technological
interests.  The $\rug$ thin films show a large THG magnitude
extending to $10^{-11}$ esu also [5].  Therefore, the fullerene
family is generally attractive for possible application to
nonlinear optical devices in near future.

In order to analyze the optical properties and to clarify mechanisms
of the large nonlinearity, we have studied the linear absorption and
the THG of $\soc$ by using a tight binding model [6] and a model
with a long-range Coulomb interaction [7].  A free electron model
yields the THG magnitudes which are in agreement with the
experiment of $\soc$ [2], while the calculated linear absorption
spectrum is not in satisfactory agreement with experiments [6].
If Coulomb interactions are taken
into account, the absorption spectra are in overall agreement
with the experiment, although the magnitude of the THG decreases
by a factor about 0.1 [7].

The first purpose of this paper is to report the details of the analysis
of optical absorption spectra of $\soc$.   In the previous brief paper [7],
the effects of the long ranged Coulomb interactions have been considered
in the tight binding model for a single molecule.  The model has been
analyzed by the restricted Hartree-Fock approximation followed by
the single excitation configuration interaction method (single CI) [8].
We use the same method in this paper.  In the experiments of actual
materials, the optical spectra become broad mainly due to the lattice
fluctuations.  We are going to simulate the effects by using the
bond disorder model with Gaussian distribution.  We will take
a sample average over disorder potentials. This method has been
used in order to describe the optical properties of conjugated
polymers [9,10].  We will show that main features observed in
experiments of $\soc$ molecules in a solution [2] are reasonably described by
the present theory.   Speaking in detail, peak positions in the frequency
dispersion of the spectra agree well with the experiment.
The relative oscillator strengths are in overall agreement with
the experimental data.  Furthermore, the broadening of absorption
peaks by lattice fluctuations is well simulated by the bond disorder
model with a reasonable disorder strength.

Secondly, we extend the calculations to one of higher
fullerenes: $\rug$.  The molecular shape becomes elongated in going
from $\soc$ to $\rug$.  Then, we expect that the optical response
varies depending on the molecular orientation relative to
the electric field of light.  We will discuss the anisotropy
for the free electron model and for the model with
the Coulomb potentials.  It is found that the optical gap is
larger when the electric field is parallel to the long axis
of the $\rug$ molecule.  The line shape of the absorption spectrum
is anisotropic, too.  These properties do not depend on whether the Coulomb
interactions are present or not.  We expect that the anisotropy
will be observed in the solids of $\rug$.

Finally, we consider a situation in which the $\rug$ molecule is
rotated randomly and the anisotropy of the optical
absorption is averaged out.  The optical spectrum
calculated by the model with Coulomb
interactions and disorder potentials are compared with the experimental
data [1] for molecules in a solution.  We will conclude that
there is an overall agreement.
The absorption around 2.7 eV in the experiment can be interpreted
as an allowed transition.
There are several allowed transitions below the energy 2.7eV,
but their oscillator strengths are smaller.  We can relate
the spectra of $\soc$ and $\rug$ by considering the symmetry
reduction from a soccerball to a rugbyball.

We shall report calculations in the following way.  In the next
section, our model and its meaning are explained.  The calculation
method of the single CI is also summarized.  In Sec. III, we report
about $\soc$ and compare with the experiment.  In Sec. IV, we
describe optical spectra of $\rug$.  The paper is closed with several remarks
in the final section.

\section{MODEL AND FORMALISM}

In order to consider optical spectra of $\soc$ and $\rug$,
we use the following hamiltonian:
\beeq
H = H_0 + H_{\rm bond} + H_{\rm int}.
\eneq
The first term $H_0$ of Eq. (2.1) is the tight binding model.
We use the following form for $\soc$:
\beeq
H_0 = {\sum_{\langle i,j \rangle, \sigma}}^{\rm D} (-t_{\rm D})
(c_{i,\sigma}^\dagger c_{j,\sigma} + {\rm h.c.})
+ {\sum_{\langle i,j \rangle,\sigma}}^{\rm S} (-t_{\rm S})
(c_{i,\sigma}^\dagger c_{j,\sigma} + {\rm h.c.}),
\eneq
where $c_{i,\sigma}$ is an annihilation operator of a $\pi$-electron at
site $i$ with spin $\sigma$; the sum with the symbol D (or S) is taken over
all the pairs $\langle i,j \rangle$ of neighboring atoms with
a double (or single) bond; and $t_{\rm D} = t + (2/3) t'$
and $t_{\rm S} = t - (1/3) t'$ are the hopping integrals.
The mean hopping is taken as $t$.  In this paper, we use the
difference between $t_{\rm D}$ and
$t_{\rm S}$, $t' = 0.1 t$, as a typical value.
The choice of $t'$ does not affect on the results so strongly,
because main contributions come from the strong Coulomb potential.
We use the next form for $\rug$:
\beeq
H_0 = - t \sum_{\langle i,j \rangle, \sigma}
c_{i,\sigma}^\dagger c_{j,\sigma},
\eneq
assuming a constant hopping integral.  We could use a realistic
combination of hoppings, but the results do not change so much
like in the $\soc$ case.  Effects of zero point vibrations and
thermal fluctuation of the lattice are modeled
by the bond disorder model which is the second term of Eq. (2.1):
\beeq
H_{\rm bond} = \sum_{\langle i,j \rangle, \sigma} \delta t_{i,j}
c_{i,\sigma}^\dagger c_{j,\sigma}.
\eneq
Here, $\delta t_{i,j}$ is the disorder potential at the bond
$\langle i,j \rangle$.   We can estimate the strength of the
disorder (standard deviation) $t_s$ from the results [11] by the
extended Su-Schrieffer-Heeger (SSH) model [12].  The value would
be $t_s \sim 0.05 - 0.1 t$.  This is of the similar magnitude
as in the fullerene tubules and conjugated polymers [13].  We shall
treat interactions among $\pi$-electrons by the following model:
\beeqa
H_{\rm int} &=& U \sum_i
(c_{i,\uparrow}^\dagger c_{i,\uparrow} - \frac{1}{2})
(c_{i,\downarrow}^\dagger c_{i,\downarrow} - \frac{1}{2})\\ \nonumber
&+& \sum_{(i,j), i \neq j} W(r_{i,j})
(\sum_\sigma c_{i,\sigma}^\dagger c_{i,\sigma} - 1)
(\sum_\tau c_{j,\tau}^\dagger c_{j,\tau} - 1),
\eneqa
where $r_{i,j}$ is the distance between the $i$th and $j$th sites and
\beeq
W(r) = \frac{1}{\sqrt{(1/U)^2 + (r/r_0 V)^2}}
\eneq
is the Ohno potential.  The quantity $U$ is the strength of
the onsite interaction, $V$ means the strength of the long range
Coulomb interaction, and $r_0$ is the average bond length.

The model is treated by the Hartree-Fock approximation
and the single CI [8].  After the Hartree-Fock approximation
$H \Rightarrow H_{\rm HF}$, we divide the total hamiltonian
as $H = H_{\rm HF} + H'$.  The term $H'$ becomes
\beeqa
H' &=& U \sum_i
(c_{i,\uparrow}^\dagger c_{i,\uparrow} - \rho_{i,\uparrow})
(c_{i,\downarrow}^\dagger c_{i,\downarrow} - \rho_{i,\downarrow})\\ \nonumber
&+& \sum_{(i,j),i \neq j} W(r_{i,j})
[\sum_{\sigma,\tau} ( c_{i,\sigma}^\dagger c_{i,\sigma} - \rho_{i,\sigma})
(c_{j,\tau}^\dagger c_{j,\tau} - \rho_{j,\tau})\\ \nonumber
& &+ \sum_\sigma ( \tau_{i,j,\sigma} c_{j,\sigma}^\dagger c_{i,\sigma}
+ \tau_{j,i,\sigma} c_{i,\sigma}^\dagger c_{j,\sigma}
- \tau_{i,j,\sigma} \tau_{j,i,\sigma})],
\eneqa
where $\rho_{i,\sigma} = \langle c_{i,\sigma}^\dagger c_{i,\sigma} \rangle$
and $\tau_{i,j,\sigma} = \langle c_{i,\sigma}^\dagger c_{j,\sigma} \rangle$.
When we write the Hartree-Fock ground state  $| g \rangle
= \prod_{\lambda {\rm: occupied}} c_{\lambda,\uparrow}^\dagger
c_{\lambda,\downarrow}^\dagger |0 \rangle$ and the single
electron hole excitations $|\mu \lambda \rangle = c_{\mu, \sigma}^\dagger
c_{\lambda, \tau} | g \rangle$ ($\mu$ means an unoccupied state;
we assume both singlet and triplet excitations in this abbreviated
notation), the matrix elements of the Hartree-Fock part and the excitation
hamiltonian become as follows:
\beeqa
\langle \mu' \lambda' | ( H_{\rm HF} - \langle H_{\rm HF} \rangle)
| \mu \lambda \rangle &=& \delta_{\mu',\mu} \delta_{\lambda',\lambda}
(E_\mu - E_\lambda),\\
\langle \mu' \lambda' | ( H' - \langle H' \rangle) | \mu \lambda \rangle
&=& 2J \delta_S - K,
\eneqa
where $E_\mu$ is the energy of the Hartree-Fock state,
$\delta_S = 1$ for spin singlet, $\delta_S = 0$ for spin triplet, and
\beeqa
J(\mu',\lambda';\mu,\lambda) &=& \sum_{i,j} V_{i,j}
\langle \mu' | i \rangle \langle \lambda' | i \rangle
\langle j | \mu \rangle \langle j | \lambda \rangle \\
K(\mu',\lambda';\mu,\lambda) &=& \sum_{i,j} V_{i,j}
\langle \mu' | j \rangle \langle \lambda' | i \rangle
\langle j | \mu \rangle \langle i | \lambda \rangle
\eneqa
with $V_{i,i} = U$, $V_{i,j} = W(r_{i,j})$ for $i \neq j$.
The diagonalization of the total hamiltonian $H$
gives the set of the excited states $\{ | \kappa \rangle \}$
within the single CI method.  In the actual calculation,
we limit the spin configurations to the singlet excitations.

The absorption spectrum in the case of the electric field
parallel to the $x$ axis is proportional to
\beeq
\sum_\kappa E_{\kappa}
P (\omega - E_{\kappa}) \langle g | x|\kappa \rangle
\langle \kappa | x | g \rangle.
\eneq
Here, $P (\omega) = \gamma/[ \pi (\omega^2 + \gamma^2)]$
is the Lorentzian distribution ($\gamma$ is the width),
$E_{\kappa}$ is the electron-hole excitation energy, and
$| g \rangle$ means the ground state.

In this paper, all the quantities with the energy dimension are shown
in the units of $t$.  We varied the parameters of Coulomb interaction
within $0 \leq V \leq U \leq 5t$, and we report here the two representative
cases: $U = V = 0$ and $U = 2V = 4t$.  Results in the latter case
turn out to be in overall agreement with experiments.

\section{OPTICAL ABSORPTION IN C$_{\rm {\bf 60}}$}

This section is devoted to optical spectra of $\soc$,
the structure of which is shown in Fig. 1(a).
The calculated spectra are nearly isotropic independent
of the direction of the electric field against
the molecular orientation.

\subsection{Coulomb interaction effects}

The dispersion of the optical absorption is calculated for the
free electron case $U = V = 0$, and displayed in Fig. 2(a).  The
Lorentzian broadening $\gamma = 0.01 t$ is used.  The distribution
of the relative oscillator strength and the positions of the peaks
agree with those in the previous calculations [6].  The second peak at
about $1.2t$ has the largest oscillator strength.

Fig. 2(b) shows the absorption for $U = 4 t$ and $V = 2 t$.
There are several peaks, among which the peak at $\sim 2.6 t$ has
the largest oscillator strength. This is in contrast to the free
electron case ($U=V=0$), where the oscillator strength is the
largest at around $~ 1.2 t$.  The Coulomb interaction tends to
reduce the oscillator strengths of the lower-energy peaks and
to enhance those of the higher-energy peaks.  In addition, it
shifts overall peak positions to higher energies.  Qualitatively
similar results have been reported in terms of the random phase
approximation [14].  The overall distribution of the oscillator
strengths in Fig. 2(b) is in good agreement with experiments [2] when
we assume $t=1.8$eV.  The origin of the large broadening of the
peaks may be lattice fluctuations.  This is to be discussed in
the next subsection.

Now, we can estimate a dielectric constant $\eps$.  The magnitude
of the Coulomb interaction at the mean carbon bond length is
$W(r_0) = U V /\sqrt{U^2 + V^2}$.  This gives an estimate:
$\eps = e^2 /[r_0 W(r_0)] = 3.11$.  The long distance limit
$W(r \gg r_0) \sim V r_0 / r$ gives $\eps = 2.80$.  The
independent study of the lifetime of positrons in $\soc$ gives
the high frequency dielectric constant $\eps_\infty \sim 3.5$ [15].
Therefore, the present estimation of the dielectric constant
near the value 3 would be reasonable.

\subsection{Lattice fluctuation effects}

Here, we should like to take into account of the lattice fluctuation
effects.  We simulate the effects by the bond disorder
model with a Gaussian distribution.  The application of the model
has been successful in describing the lattice fluctuation effects in
conjugated polymers [9].  The estimation by the results of the extended
SSH model for $\soc$ gives the order $t_s \sim 0.05 - 0.1 t$ [11].
We change the disorder strength around these values.  Averaging over 100
samples turned out to be enough to obtain the smooth optical spectra.

Fig. 2(c) shows the calculated optical spectrum for $t_s = 0.09t$.
For comparison we show in Fig. 2(d) the experimental data
obtained by Ren et al. for $\soc$ in a solution
and a $\soc$ thin film [2].  The data are
shown for comparison assuming $t = 1.8$eV.
Due to the broadening, the peaks near the energy $1.8t$ in Fig. 2(b)
merge into a single broad feature in Fig. 2(c).  The several peaks
around $3.4t$ obtain the same effect as well.  Overally, there are
three main features in the energy dependence of the calculated
data.  Their positions in the energy agree well with the experimental
data, $~2.0t$, $~2.6 t$, and $~ 3.1t$, obtained from the solution.
The relative oscillator strengths fairly agree also.  Furthermore,
the widths of the features can be well simulated by the broadening
from the bond disorder model.  Experimentally, the peaks would be
broadened by various kinds of phonon modes: intermolecular libration
and intramolecular vibrations.  They give rise to many sidebands,
which broaden into several main features.  The disorder strength,
adopted here, simulates well the total broadening as the net
contribution from the various phonon modes.

In the solid, the broadening is larger and a broad hump appears
around the energy 2.8eV $(=1.5t)$, as is shown in Fig. 2(d).
There is a tiny structure at 3eV in the corresponding region of
the absorption spectrum of $\soc$ in a solution (Fig. 2(d)
and Ref. 17).  We assume that the forbidden
transitions in that energy region become partially allowed due
to lattice fluctuations or intermolecular interactions.
If the lattice fluctuations are effective,
the effect can be simulated by the bond disorder model.
In Fig. 2(c), the absorption in the low energy part
multiplied by the factor 10 is also shown.  The several forbidden
transitions around $1.2t - 1.6t$ become allowed by the disorder,
giving rise to an absorption tail in the lower energy region.
The relative magnitude is similar to the solution data
shown in Fig. 2(d).  This might be also the origin of the 2.8eV
hump in the solid.  However, the strength of the absorption relative to the
main peaks in our theory is about one order of magnitude smaller than
that in the experiment.  If we increase disorder drastically
to explain the magnitude, we get absorption peaks that are too
broad compared with the experiments.  This is true for all the
cases of Coulomb interaction parameters we tried.  Therefore, it seems
difficult to explain the oscillator strength of the partially
allowed transitions in solids quantitatively.  Additional effects
such as intermolecular interactions will be necessary, and this
problem will be investigated in a separate paper [16].
We note that the $\soc$ oxide shows a weak absorption peak
in a similar energy region [18].  This could also
be interpreted by using a model with impurities.

\section{OPTICAL ABSORPTION IN C$_{\rm {\bf 70}}$}
\subsection{Anisotropy effects}

The $\rug$ molecule has a rugbyball-like shape.  Figs. 1(b) and (c)
display the molecule.  We locate the long axis of the molecule
along the $x$ direction.  Fig. 1(b) shows the molecule which
is projected along the $z$ axis. The $x$ axis goes from the left
to the right of the figure.  Fig. 1(c) is the molecule projected
along the $x$ axis.  We expect that the absorption becomes
anisotropic with respect to the direction of the electric field of light.
We look at anisotropy effects in the free electron model as well
as in the model with Coulomb interactions.

Figs. 3 (a), (b), and (c) show absorption spectra of the free electron
case when the electric field is in the $x$, $y$, and $z$
directions, respectively.  Figs. 3 (b) and (c) have the almost same
spectral shape.  But, the spectral shape is very different when the
electric field is parallel to the long axis.   The optical gap is about 0.7$t$
in Fig. 3(a), while it is about 0.6$t$ in Fig. 3(b) and (c).  This accords
with the result of the recent H\"{u}ckel calculation [19] in that the optical
gap of $\rug$ is larger when the electric field is parallel to  the
long axis of the molecule.

Next, we shall turn to Coulomb interaction effects.  We display the
optical spectra for $U = 4t$ and $V = 2t$.  We will later find that these
parameters are appropriate for $\rug$ also.  Figs. 4 (a), (b), and (c)
show the spectra for the respective molecular
orientations corresponding to Figs. 3(a-c).  It seems that there is
a strong anisotropy when the Coulomb interactions are present, too.
The optical gap in Fig. 4(a) is about $1.1t$ with a weak
absorption peak.  It is about $1.0t$ in Figs. 4 (b) and (c).
Therefore,  the optical gap becomes larger when the electric field
is parallel to the long axis of the molecule.  This property
does not change whether there are Coulomb interactions or not.
We highly expect that the anisotropy will be clearly observed in the solid
of $\rug$.

\subsection{Orientationally Averaged Spectra}
\subsubsection{Coulomb interaction effects}

The optical absorption data have been reported for molecules
in solution [1].  In order to compare the present theory
with the experiment, we have rotated the molecule randomly
and averaged the spectra over 100 orientations.  Averaged data without
bond disorder are shown in Figs. 5(a) and (b).  We use the
Lorentzian broadening $\gamma = 0.01 t$ as before.  Fig. 5(a) is
for the free electron case, and Fig. 5(b) is for the Coulomb
interaction strengths $U= 4t$ and $V = 2t$.  We can easily
see that the peaks in three figures of Fig. 3 remain in Fig. 5(a).
The same thing can be said in the relation between Fig. 4
and Fig. 5(b).  The effects of the Coulomb interactions are
similar to those of the $\soc$ case: the Coulomb potentials
tend to shift peaks to higher energies, and oscillator strengths
at higher energies have relatively larger weight than in
the free electron case.  The shift is due to the fact that the
inter site Coulomb interactions give rise to bond order
parameters in the Hartree-Fock approximation.  The bond order
effectively enhances the intersite hopping integrals,
and thus the energy gap.

\subsubsection{Lattice fluctuation effects}

Now, we treat lattice fluctuation effects
by using the bond disorder model.  As we expect that the width of
lattice fluctuations in $\rug$ is similar to that in
$\soc$ even if the molecular structure is different, we assume
the same $t_s = 0.09t$ in this calculation again.  The orientation
of the molecule and disorder potentials are changed 100 times,
and the optical spectra are averaged.

Fig. 5(c) shows the calculated optical absorption.  Fig. 5(d) shows
the experimental data of $\rug$ in a solution taken from [1].  The
abscissa is scaled by using $t = 1.9$eV.  The optical gap near $1.0 t$
accords well with the experiment.  The small peaks around $1.8t$
may correspond to those of the experiment in the similar energy
region.  Commonly, there is a dip of the optical
absorption at about $2.0t$ in the calculation and experiment.
There is a maximum at $2.8t$ in the two data.  Therefore, we
conclude that there is an overall agreement between the calculation
and the experiment.  The quantum lattice fluctuation effects are
also simulated well by the bond disorder model.

\section{CONCLUDING REMARKS}

The most striking difference between optical spectra of $\soc$ in
solution and of $\soc$ solids is that the feature around 2.8eV has the
strong oscillator strength only in solids.  In the present
calculation for $U = 4t$ and $V = 2t$ without disorder, there
are dipole forbidden transitions at the corresponding energy.
The disorder makes these transitions partially allowed.
We have discussed that this effect is one of the origins
which make the forbidden transitions observed.  However,
the theoretical oscillator strengths seem to be smaller
by an order of magnitude than that in the experiments.  We
need further study in order to solve this problem.  For example,
we are now studying intermolecular interaction effects
in addition to lattice fluctuations.  This will be reported separately [16].

The absorption spectra of $\rug$ become highly anisotropic
depending on the orientation of the molecule relative to the polarization
of light.  Now, single crystals of $\rug$ have been made
and their optical data have been reported [5].  Therefore, an extension
of the present calculations to $\rug$ solids is quite interesting.
We expect that the anisotropy of the optical spectra of the molecule
will remain in solids, too.

In view of the agreement of our theory  with experiments in $\soc$
and $\rug$, it is very interesting to look at optical properties and
excitonic effects of higher fullerenes where the number of carbons is
larger than 70.  The method of the single CI should be valid
enough in these systems, too.  The number of isomers
rapidly increases as the carbon number increases as 76, 78,
82, 84, and so on.  The calculated optical data will reflect
the electronic and lattice structures of isomers, and
they will be mutually different.  Thus, the calculations will be
useful in order to distinguish isomers by using optical
experiments.

\mbox{}

\noindent
{\bf ACKNOWLEDGEMENTS}\\
The authors acknowledge helpful correspondence with Dr. Mitsutaka
Fujita.  They also thank Mr. Mitsuho Yoshida for providing them
with the molecular coordinate of $\rug$ which is generated by
the program [20] based on the projection method on the triangular [20]
and honeycomb lattices [21].  A part of the work was done while
one of the authors (K.H.) was staying at the University of Sheffield,
United Kingdom.  He acknowledge the hospitality that he obtained
from the University.

\pagebreak
\begin{flushleft}
{\bf REFERENCES}
\end{flushleft}

\noindent
* Electronic mail address: harigaya@etl.go.jp.\\
$[1]$ J. P. Hare, H. W. Kroto, and R. Taylor, Chem. Phys. Lett.
{\bf 177}, 394 (1991).\\
$[2]$ S. L. Ren, Y. Wang, A. M. Rao, E. McRae, J. M. Holden, T. Hager,
KaiAn Wang, W. T. Lee, H. F. Ni, J. Selegue, and P. C. Eklund,
Appl. Phys. Lett. {\bf 59}, 2678 (1991).\\
$[3]$ J. S. Meth, H. Vanherzeele, and Y. Wang, Chem. Phys. Lett.
{\bf 197}, 26 (1992).\\
$[4]$ Z. H. Kafafi, J. R. Lindle, R. G. S. Pong, F. J. Bartoli,
L. J. Lingg, and J. Milliken, Chem. Phys. Lett. {\bf 188},
492 (1992).\\
$[5]$ F. Kajzar, C. Taliani, R. Danieli, S. Rossini, and
R. Zamboni, (preprint).\\
$[6]$ K. Harigaya and S. Abe, Jpn. J. Appl. Phys. {\bf 31},
L887 (1992).\\
$[7]$ K. Harigaya and S. Abe, J. Lumin. (to be published).\\
$[8]$ S. Abe, J. Yu, and W. P. Su, Phys. Rev. B {\bf 45},
8264 (1992).\\
$[9]$ S. Abe, M. Schreiber, W. P. Su, and J. Lu,
Mol. Cryst. Liq. Cryst. {\bf 217}, 1 (1992).\\
$[10]$ K. Harigaya, A. Terai, Y. Wada, and K. Fesser,
Phys. Rev. B {\bf 43}, 4141 (1991).\\
$[11]$ B. Friedman and K. Harigaya, Phys. Rev. B {\bf 47},
3975 (1993); K. Harigaya, Phys. Rev. B {\bf 48}, 2765 (1993).\\
$[12]$ W. P. Su, J. R. Schrieffer, and A. J. Heeger, Phys. Rev. B
{\bf 22} 2099 (1980).\\
$[13]$ R. H. McKenzie and J. W. Wilkins, Phys. Rev. Lett. {\bf 69},
1085 (1992).\\
$[14]$ G. F. Bertsch, A. Bulgac, D. Tom\'{a}nek, and Y. Wang,
Phys. Rev. Lett. {\bf 67}, 2690 (1991);  Y. Wang, G. F. Bertsch,
and D. Tom\'{a}nek, Z. Phys. D {\bf 25}, 181 (1993).\\
$[15]$ S. Ishibashi, N. Terada, M. Tokumoto, N. Kinoshita, and
H. Ihara, J. Phys.: Condens. Matter {\bf 4}, L169 (1992).\\
$[16]$ S. Abe and K. Harigaya, (in preparation).\\
$[17]$ Y. Wang and L. T. Cheng, J. Phys. Chem. {\bf 96},
1530 (1992).\\
$[18]$ K. M. Creegan, J. L. Robbins, W. K. Robbins, J. M. Millar,
R. D. Sherwood, P. J. Tindall, and D. M. Cox, J. Am. Chem. Soc.
{\bf 114}, 1103 (1992).\\
$[19]$ J. Shumway and S. Satpathy, Chem. Phys. Lett. {\bf 211},
595 (1993).\\
$[20]$ M. Yoshida and E. Osawa, (preprint); M. Yoshida and E. Osawa,
the Japan Chemistry Program Exchange, Program No. 74.\\
$[21]$ M. Fujita, R. Saito, G. Dresselhaus, and M. S. Dresselhaus,
Phys. Rev. B {\bf 45}, 13834 (1992).\\

\pagebreak
\begin{flushleft}
{\bf FIGURE CAPTIONS}
\end{flushleft}

\noindent
FIG. 1.  Molecular structures for (a) $\soc$, and (b,c) $\rug$.
In (b), the long axis ($x$ axis) penetrates from the left to the right
of the figure.  The mirror symmetry planes are perpendicular to the
figure.  In (c), the $\rug$ is projected along the long axis.

\mbox{}

\noindent
FIG. 2.  Optical absorption spectra for $\soc$ shown in
arbitrary units.  The abscissa is scaled by $t$.
The spectra are calculated with the parameters (a) $U = V = 0$,
(b,c) $U = 4t$ and $V = 2t$.  The Lorentzian broadening
$\gamma = 0.01 t$ is used.  In (c), lattice fluctuations
are taken into account by the bond disorder of the strength
$t_s = 0.09t$.  (d) The experimental spectra (Ref. [2]) of molecules
in a solution (solid line) and of a $\soc$ film (dotted line).
We use $t = 1.8$eV.

\mbox{}

\noindent
FIG. 3.  Optical absorption spectra for $\rug$ shown
in arbitrary units.  The abscissa is scaled by $t$.
The interaction strengths are $U = V = 0$.
The Lorentzian broadening $\gamma = 0.01 t$ is used.
In (a), (b), and (c), the electric field is parallel to
the $x$, $y$, and $z$ axes, respectively.

\mbox{}

\noindent
FIG. 4.  Optical absorption spectra for $\rug$ shown
in arbitrary units.  The abscissa is scaled by $t$.
The interaction strengths are $U = 4t$ and $V = 2t$.
The Lorentzian broadening $\gamma = 0.01 t$ is used.
In (a), (b), and (c), the electric field is parallel to
the $x$, $y$, and $z$ axes, respectively.

\mbox{}

\noindent
FIG. 5.  Orientationally averaged  optical absorption spectra
for $\rug$ shown in arbitrary units.  The abscissa is scaled by $t$.
The spectra are calculated with the parameters (a) $U = V = 0$,
(b,c) $U = 4t$ and $V = 2t$.  The molecule is rotated randomly
in order to average the anisotropy.  The Lorentzian broadening
$\gamma = 0.01 t$ is used.  In (c), lattice fluctuations
are taken into account by the bond disorder of the strength
$t_s = 0.09t$.  (d) The experimental spectrum (Ref. [1]) of
molecules in a solution.  We use $t = 1.9$eV.


\end{document}